# Observation of an Abrupt 3D-2D Morphological Transition in Thin Al Layers Grown by MBE on InGaAs surface

A. Elbaroudy,[1,2 a)] B. Khromets,[3,4] F. Sfigakis,[3,4] E. Bergeron,[3,4] Y. Shi,[5] M.C.A. Tam,[1] T. Blaikie,[1] J. Baugh,[2,3,4] and Z. R. Wasilewski[1,2,3,5]

1 Department of Electrical and Computer Engineering, University of Waterloo, Waterloo N2L 3G1, Canada
2 Waterloo Institute for Nanotechnology, University of Waterloo, Waterloo N2L 3G1, Canada
3 Institute for Quantum Computing, University of Waterloo, Waterloo N2L 3G1, Canada
4 Department of Chemistry, University of Waterloo, Waterloo N2L 3G1, Canada
5 Department of Physics, University of Waterloo, Waterloo N2L 3G1, Canada

a) Electronic mail: ammaelbaroudy@uwaterloo.ca

Among superconductor/semiconductor hybrid structures, in-situ aluminum (Al) grown on InGaAs/InAs is widely pursued for the experimental realization of Majorana Zero Mode quasiparticles. This is due to the high carrier mobility, low effective mass, and large Landé g-factor of InAs, coupled with the relatively high value of the in-plane critical magnetic field in thin Al films. However, growing a thin, continuous Al layer using the Molecular Beam Epitaxy (MBE) is challenging due to aluminum's high surface mobility and tendency for 3D nucleation on semiconductor surfaces. A study of epitaxial Al thin film growth on $In_{0.75}Ga_{0.25}As$ with MBE is presented, focusing on the effects of the Al growth rate and substrate temperature on the nucleation of Al layers. We find that for low deposition rates, 0.1 Å/s and 0.5 Å/s, the growth continues in 3D mode during the deposition of the nominal 100 Å of Al, resulting in isolated Al islands. However, for growth rates of 1.5 Å/s and above, the 3D growth mode quickly transitions into island coalescence, leading to a uniform 2D Al layer. Moreover, this transition is very abrupt, happening over an Al flux increase of less than 1%. We discuss the growth mechanisms explaining these observations.   The results give new insights into the kinetics of Al deposition and show that with sufficiently high Al flux, a 2D growth on substrates at close to room temperature can be achieved already within the first few Al monolayers. This eliminates the need for complex cryogenic substrate cooling and paves the way for the development of high-quality superconductor-semiconductor interfaces in standard MBE systems.

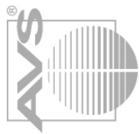

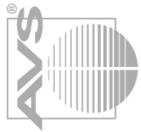

# I. INTRODUCTION

The combination of superconductor (SP) and semiconductor (SE) heterostructures has garnered significant attention in the field of condensed matter physics. In particular, these structures are crucial to the progression of topological quantum computation through braiding-based techniques[1,2] and transmon qubits[3]. The pursuit of topological qubits within SP/SE material systems relies on two primary methods: one-dimensional self-assembled nanowires[4] and gated two-dimensional electron gas (2DEG) systems[5]. For the latter approach, Al deposited in situ on a quantum well surface made of InAs/InGaAs is the most promising platform. Al thin films ($\sim$10 nm thick) are especially well suited for the superconductive layer due to their high values of superconducting coherence length, which is the measure of the size of a Cooper pair (distance between the two electrons), and relatively high value of the in-plane critical magnetic field[5,6]. InAs is a preferred material choice for the quantum well due to its large Landé g-factor, strong spin-orbit interaction, and small effective mass, meeting the conditions for the formation of Majorana zero modes (MZMs)[7,8]. Majorana quasiparticles offer a promising platform for robust practical qubits due to their non-Abelian statistics and potential for topological protection[7]. Self-assembled nanowires have been successfully used in experimental studies[9,10], yet scaling up the nanowires to large device arrays for the Majorana quasiparticle braiding or fusion applications would be very challenging. On the other hand, top-down fabrication techniques for 2DEG system design are more promising for practical applications and large-scale production[11].

Sub-gap states are typical in the materials with ex-situ Al deposition[12]. Specifically, defects due to impurities at the metal-semiconductor interfaces enhance the density of sub-gap states, resulting in undesirable dissipative effects. These sub-gap states reduce the qubit coherence time by randomly absorbing energy and adding noise to the stored phase of a qubit state[13]. Therefore, an atomically clean metal-semiconductor interface is essential for the development of topological qubits. In-situ Al deposition in the MBE system, following the growth of the InAs quantum well, yields an improved SP/SE interface of the hybrid heterostructure materials system, resulting in a hard superconducting gap with no sub-gap states[11,12]. Aside from purity, essential to viable devices are very thin, flat, and continuous Al layers. These have been consistently demonstrated only in systems capable of Al deposition at cryogenic temperatures[14], while broadly accessible MBE growth at temperatures closer to 300 K typically results in a distinct 3D island morphology. Several fundamental studies have focused on the in-situ epitaxial deposition of Al on a variety of semiconductor surfaces. This includes Al on GaAs[15,16], InGaAs[5,15], and more recently, 2DEG systems utilizing near-surface InAs quantum wells[11].

In the work done by Pilkington[15], a consistent trend was observed for growing 1200 Å thick epitaxial Al on arsenide semiconductors at room temperature with growth rates of 1 Å/min and 20 Å/min. Initially, an adlayer formed on the semiconductor surface, followed by 3D nucleation in three orientations: (1 0 0), (1 1 0), and (1 1 0)R. The islands that formed on these nucleation centers continued to grow as more Al was deposited. At layer thicknesses above about 400 Å, the (1 0 0) orientation began to dominate, transitioning the growth into a partial 2D mode. This process was assessed by Reflection High-Energy Electron Diffraction (RHEED) analysis and Atomic Force Microscopy (AFM) imaging. The latter revealed that the final surface was not that of a continuous flat layer but had a morphology characterized by hillock-valley structures, regardless of the initial buffer layer compositions. A more recent study[17] reported that when depositing Al at thicknesses of 35.9 Å and 71.8 Å on InAs, at temperatures below 0 °C and with a growth rate of about 0.09 Å/s, Al tends to form 3D islands rather than a 2D smooth layer. It was observed that on InAs, the height of these islands was reduced when grown on thicker AlAs intermediate layers, suggesting the potential for achieving a flat Al film by adjusting this layer's thickness. The study found that Al films grown on both AlAs and $In_{0.91}Ga_{0.09}As$ did not exhibit laminar growth either, forming 3D islands instead. Additionally, Ga-rich surfaces showed significantly less reactivity with Al compared to InAs or AlAs surfaces. Recently, Cheah et al.[11] investigated the growth of 120 Å of Al on a 2DEG system consisting of an InAs quantum well with a 13.4 nm top barrier of $In_{0.75}Ga_{0.25}As$, using an Al growth rate of 1 Å/s and a nominal substrate temperature of approximately -30 °C. Intermediate layers of 2 and 5 monolayers of GaAs were used to prevent indium diffusion into Al and to improve chemical stability during device fabrication. This approach was found to enhance mobility and the strength of the induced proximity effect without affecting the electron mobility or superconducting properties of the system.

Despite these different studies, a comprehensive analysis of growth parameters influencing the growth mode of Al, particularly the effects of the Al growth rate and substrate temperature during deposition, has not been reported yet. In the present work, we study the quality of epitaxial Al thin films (∼10 nm) grown by MBE directly on $In_{0.75}Ga_{0.25}As$ with no intermediate layers, at temperatures above zero °C. We discuss the influence of Al growth rates and substrate temperature on the Al nucleation path and final layer morphology. Distinct Al islands were observed at relatively low Al deposition rates. Even though 3D nucleation is observed for all cases, it transitions quickly to 2D growth mode for the fastest deposition rates, resulting in near-atomically flat final Al films, as evidenced by Scanning Electron Microscope (SEM) and AFM measurements. Moreover, we present measurements of the critical in-plane magnetic field for the superconductor to normal transition for the grown Al layers. This work sheds new light on the growth mechanisms of Al thin films, carrying implications for the advancement of topological superconducting qubits.

## II. EXPERIMENTAL

### A. Molecular Beam Epitaxy Growth

In this study, we present experimental results based on two sets of samples grown during two different MBE campaigns, A and B. The nominal structures of all samples were the same and are shown in **Figure 1**. Also, all the growth steps were the same up to and including the top $In_{0.75}Ga_{0.25}As$ layer. The key difference between the samples was in the deposition rate of Al, as detailed in **Table 1**. Also, for the samples grown during Campaign A, the main manipulator shutter was not available. As detailed later, this resulted in a higher initial substrate temperature during Al deposition.

All the samples were grown using a Veeco GEN10 MBE system. The structures were grown on quarters of 3" semi-insulating, Fe-doped, (001) InP epi-ready substrates. The wafers were In-free mounted, sandwiched between two Mo plates mounted in a wafer holder designed for 3" substrates, all supplied by Veeco. Initially, the substrates were outgassed at 200 °C for two hours in the system's load lock. This was followed by a transfer to the preparation module for an additional one hour of outgassing at 400 °C. Subsequently, the substrates were moved to the growth module to begin the epitaxial growth process. The native oxide was removed by elevating the substrate temperature to 535 °C under $As_4$ overpressure with substrate rotation at 20 rpm. The wafer's temperature was measured using band edge thermometry (BET)[18]. Following this, the substrate temperature was lowered to 470 °C for the growth of a nominally lattice-matched 100-nm-thick $In_{0.53}Ga_{0.47}As$ buffer layer, followed by a 6 nm thick layer of $In_{0.75}Ga_{0.25}As$ (**Figure 1**). The latter constitutes a typical top barrier for a shallow InAs QW used to fabricate structures supporting MZMs. The RHEED diffraction patterns along the (110) and (-110) substrate azimuths were continuously monitored on a rotating wafer using a kSA 400 RHEED acquisition system triggered at appropriate azimuths. After the growth of the $In_{0.75}Ga_{0.25}As$ layer, the resulting films exhibited a consistent and well-ordered (2x4) reconstruction, indicative of high-quality 2D growth and precise surface stoichiometry control. Upon completing the growth of the top $In_{0.75}Ga_{0.25}As$ layer, the substrate's temperature was lowered to 400 °C, at which point the As flux was interrupted by closing the As cracker valve and the As cell shutter. For series B, to reduce surface contamination and prevent substrate heating by the idling effusion cells, the main manipulator shutter located beneath the substrate was closed. The wafer's temperature was maintained at 400 °C for a duration of at least four hours. This step was introduced to eliminate condensation of As on the substrate at lower temperatures and to ensure that the subsequent growth of epitaxial Al was taking place on a possibly clean (2×4) $In_{0.75}Ga_{0.25}As$ surface. Once the background pressure in the growth module decreased to below 1.2E-10 Torr, the manipulator heater power was reduced to zero, and the wafers were allowed to cool down overnight in the growth module prior to the Al deposition. Helped by a liquid nitrogen-cooled shroud environment, the substrates reached temperatures below 13 ℃ by the following morning for the series B samples. Because of the lack of protection from the main shutter, the lowest temperature achieved by samples from series A was between 20 and 30 °C. For the deposition of the Al layer, both Al cells installed in the reactor were used simultaneously, and Al deposition was done on stationary substrates. For each growth rate used, the Al cells were ramped up from their idle temperature of 750 ℃ to the target temperatures shortly before the Al deposition to minimize the thermal radiation load in the reactor. For samples in series B, the main shutter was opened throughout the duration of Al deposition only. After deposition, the temperatures of the Al cells were quickly ramped down back to idling.

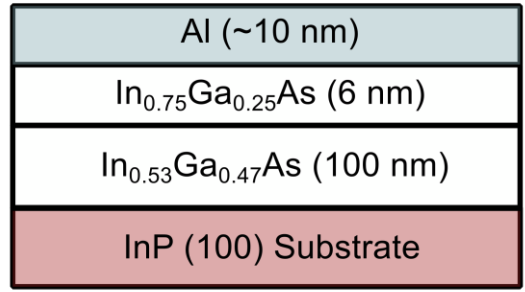

**Figure 1.** Material structures of all samples grown in this study.

**Table 1.** List of all samples reported in this study.

| Sample ID | Al growth rate (Å/s) | Al thickness (Å) |
|-----------|----------------------|------------------|
| A1 | 0.1 | – |
| A2 | 0.5 | – |
| A3 | 2.0 | ~ 100 |
| B1 | 0.1 | – |
| B2 | 0.5 | – |
| B3 | 1.5 | ~ 85 |
| B4 | 2 | ~ 100 |
| B5 | 3 | ~ 100 |

## B.  Substrate Temperature Monitoring

The temperature of the substrates was monitored throughout the entire growth process using a BET technique utilizing two array spectrometers, NIR for the 900-1700 nm spectral range and Vis-IR for the 400-1050 nm spectral range.  This allowed us to track InP absorption edge location, and thus the substrate temperature, through the entire range of temperatures accessible in our system. The NIR spectrometer was used for oxide desorption and InGaAs layers deposition, while the Vis-IR spectrometer was used to measure the substrate's temperature in the 0-150 °C range. Since the manipulator heater was turned off during the Al deposition, the radiation needed to track the substrate absorption edge was supplied by a halogen lamp located at the top of the manipulator. The light was guided to the back of the substrate by a quartz light pipe installed in the manipulator. Dedicated tests performed before the experiments revealed that the halogen power level used did not cause significant local heating of the samples prepared for Al deposition. The raw transmission spectra were subsequently used to determine the band edge position. During the growth process, the radiation emitted from the Al cells led to a gradual increase in the substrate temperature. Since the thermal radiation power emitted from the Al cell is expected to be proportional to the fourth power of the Al melt temperature, while the Al flux (thus the growth rate) increases exponentially with increasing the melt temperature, the total thermal radiation dose absorbed by the wafer during Al deposition was expected to be lowest for the fastest Al deposition rates. **Figure 2a-c** compares the evolution of substrate temperature over time during Al deposition for the three samples of series A. The manipulator thermocouple readings, $T_{thr}$, are shown in addition to the actual wafer temperature, $T_{BET}$, measured with BET. The thermocouple junction is located in the space between the heater element and the wafer and is not in contact with the wafer.

As expected, the layer deposited at the slowest growth rate – sample A1 – received the highest thermal dose and heated up by more than 90 °C during the Al deposition cycle. Remarkably, the thermocouple readings, $T_{thr}$, increased by less than 10 °C during the same period. This emphasizes the importance of monitoring the wafer temperature directly. The increase in the substrate temperature during the deposition process is recorded for all the samples reported here and those are listed in **Table 2**.

**Table 2.** Substrate temperature measurements for all samples during Al growth. This table presents '$T_{BET}$ Start', which indicates the substrate temperature at the beginning of Al growth. '$\Delta T_{BET}$' and '$\Delta T_{thr}$' represent the differences in substrate temperature between the start and end of the growth period, measured by BET and thermocouple, respectively.

| Sample ID | $T_{BET}$ start (°C) | $\Delta T_{BET}$ (°C) | $\Delta T_{thr}$ (°C) |
|---|---|---|---|
| A1 | 25.3 | 90.6 | 10.0 |
| A2 | 26.3 | 51.2 | 0.94 |
| A3 | 29.7 | 23.6 | 0.4 |
| B1 | 8.0 | 103.3 | 13.6 |
| B2 | 13.0 | 88.1 | 3.3 |
| B3 | 12.1 | 49.6 | 0.6 |
| B4 | 12.8 | 39.4 | 0.8 |
| B5 | 11.3 | 38.6 | 0.6 |

**Figure 2d** shows the evolution of $T_{BET}$ for all three samples as a function of the cumulative energy output $D(t)$ from the unit surface area of the molten aluminum in the Al cells. Here, we used the approximation:

$$D(t) = \frac{1}{2}\sigma(T_1^4 + T_2^4)$$

where $\sigma = 5.67 \cdot 10^{-8} \text{W} \cdot \text{m}^{-2} \cdot \text{K}^{-4}$ is the Stefan-Boltzmann constant for black body radiation, and we assume the emissivity of the cell to be close to one. The $T_1$ and $T_2$ are the absolute Al melt temperatures for the two aluminum cells used. Without heat losses from the wafers, we would expect all the points in **Figure 2d** to fall on the same curve and wafer temperature to remain constant after Al deposition. The approximately exponential dependencies seen for the wafer heating and cooling phases in **Figure 2a-c** are the direct measure of the heat loss channels present. Although we do expect some radiative heat loss by the wafers, most likely, the main heat loss is through the wafer's direct contact with the holder and through it to the rest of the fairly massive manipulator, whose temperature is not altered much during the Al deposition cycles. The relative displacement of the traces in **Figure 2d** may also be caused by the deviation of the effective cell orifice radiation power from the assumed $T^4$. Both Al cells used here are dual filament cells equipped with 40 cc conical PBN crucibles. Only bottom filaments are used, though, and the crucible has a strong temperature gradient from the melt to the top of the cell. Also, this gradient is expected to change significantly between the lowest and the highest Al melt temperatures used in the experiments reported in this study. None of these effects are reflected in the equation used.

Still, additional complications exist. The wafer heating is affected not only by the radiation absorbed by the wafer itself but also by the heating rate of the holders. The latter will depend on the materials deposited on them in the past. We see evidence that such quarter wafers are not uniformly heated by radiation from Al cells. Such temperature gradients most likely result from the variation in wafer-holder contact along the wafer periphery. One other element that can affect the heating rate of the substrate exposed to a fixed thermal radiation load is the light absorption in the small bandgap InGaAs epitaxial layers, and also in the interfacial InAs layer. The latter is typically formed on InP substrates during oxide desorption under As overpressure and is due to the displacement of phosphorus by arsenic. From the fitting of a dynamical scattering theory to XRD measurements, we find the thicknesses of such InAs layers varying between about 0.5 to 1.5 ML. Thus, any variation in the InGaAs compositions or InAs interfacial layer thickness may also influence the observed heating rate and the maximum temperature achieved during deposition.

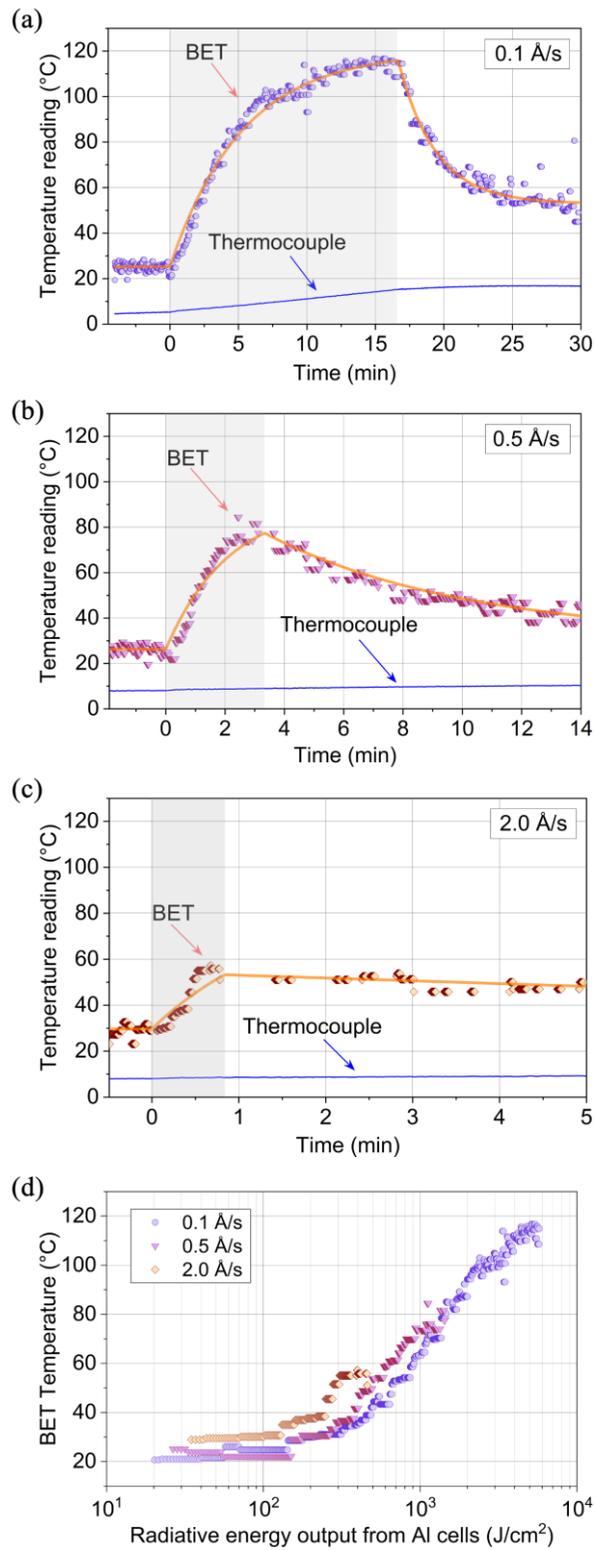

**Figure 2.** Substrate temperatures during the growth of Al layers for series A samples: (a) A1, (b) A2, and (c) A3. The shaded regions within the temperature data indicate the periods of opening the Al cells' shutters, marking the duration of Al growth. Figure (d) shows the relationship between the radiative energy output from the Al cells as a function of the progression of $T_{BET}$ for all samples in series A.

Notably, even though the slowest deposition rate leads to the largest total absorbed radiative power and results in the largest temperature increase by the wafer, the rate of temperature rise increases with increasing Al growth rate. For example, over the first 150 seconds, the temperature of sample A2 (0.5 Å/s) rose by over 40 °C, while an increase of only about 25 °C was observed for sample A1 (0.1 Å/s). We will come back to this observation in the discussion section.

# III. RESULTS AND DISCUSSION

## A. Scanning Electron Microscope (SEM)

The surface morphology of all the series B samples was tested using a JEOL JSM-7200F SEM. All scans were conducted at the same magnification for comparison purposes (**Figure 3**). The sample grown at 0.1 Å/s shows small 3D islands of Al with small separations, while the sample grown at 0.5 Å/s shows a larger radius of 3D islands of Al with larger separations, as depicted in **Figure 3(a) and (b)**, respectively. With increasing the growth rate to 2 Å/s, a continuous Al layer with occasional very small voids was observed, as seen in **Figure 3(c)**. Finally, at the highest tested growth rate of 3 Å/s, the smoothest surface was achieved, characterized by an Al layer free of voids, as seen in **Figure 3(d)**.

For the B3 sample, the Al cells' temperatures ramped down during the deposition, leading to a change in Al growth rate from 2 to 1.4 Å/s over 50 seconds and resulting in a deposition of approximately 8.5 nm thick Al film. As in the other four series B samples, this layer was deposited on a stationary substrate, which led to a variation of Al flux across the wafer. This deposition resulted in a wafer covered only partially by a continuous smooth Al layer. The remaining part showed rough morphology with distinct Al islands. As shown in **Figure 4**, the 3D-2D transition region is abrupt, happening across a fraction of a millimetre. Across this region, round 3D Al clusters transition into elongated clusters, which then merge to form a continuous Al layer. Remarkably, the change of Al flux across this transition region is less than 1% based on the measurements from B3 sample[19]. We do expect some temperature gradient across the wafer. Still, the variation in the local wafer temperature across this region is expected to be only a fraction of a degree throughout the entire deposition process. This abruptness of the observed 3D-2D morphological transition for thin Al layers here makes it akin to other phase transitions, such as melting points of solids.

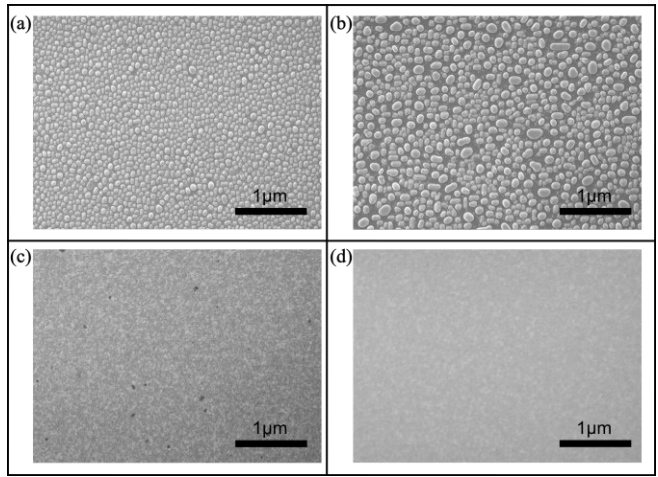

**Figure 3.** High-resolution Scanning Electron Microscope images illustrating the variations in surface morphology among series B samples (a) B1, (b) B2, (c) B3, and (d) B4.

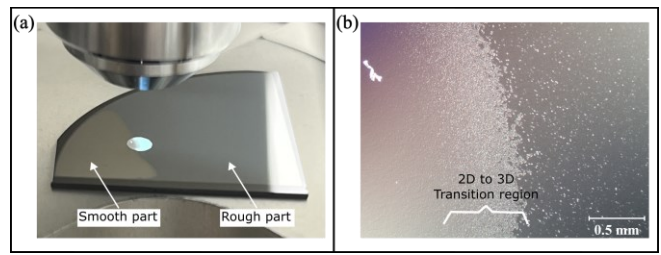

**Figure 4.** (a) B3 Sample Surface: Shows the rough and smooth areas, highlighting the light spot analyzed in (b). Figure (b) Nomarski Microscopy image illustrates the transition from continuous to discontinuous Al layers.

## B.   Atomic Force Microscopy

To gain further insight into the series B samples' morphologies, we performed atomic force microscopy measurements. These were done using a Bruker Dimension FastScan AFM system. **Figure 5** illustrates the shift in nucleation behavior of the Al thin layers, marking a transition from 3D Al islands formed at lower growth rates, as shown in **Figure 5(a)** and **(b)**, to a 2D laminar Al layer, shown in **Figure 5(c)** and **(d),** with the increase in deposition rate from 0.1 Å/s to 3 Å/s. To better illustrate the quality of the continuous B4 and B5 layers, the cross sections of these layers are also shown in panels **5(c)** and **5(d)**. This was derived from the surface morphology scan along the lines between the arrows.

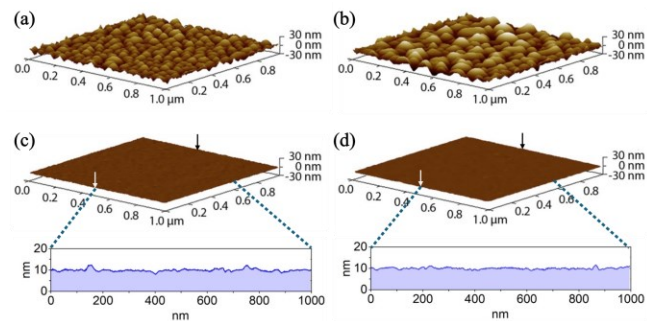

**Figure 5.** Atomic Force Microscopy illustrates the variations in surface morphology among samples: (a) B1, (b) B2, (c) B4, and (d) B5. The root mean square roughness for the B1, B2, B3 and B4 samples are 5.68 nm, 9.2 nm, 0.549 nm and 0.385 nm, respectively.

### C.  Low-temperature resistivity measurements with in-plane magnetic field

Using standard AC lock-in techniques, the transition from superconducting to normal resistance in our epitaxial Al films in series B samples was characterized as a function of the magnetic field. These experiments were conducted in a helium-3/helium-4 dilution refrigerator (model TLM from Oxford Instruments) equipped with a 20 Tesla magnet and achieving a base temperature of 12 mK. Four-terminal measurements (I+, I-, V+, V-) on cleaved small samples (3 mm x 1 mm) allowed for the precise determination of the epitaxial Al resistance, effectively eliminating the influence of contact resistance, line resistances, and filters, among other factors. **Figure 6** illustrates the critical in-plane magnetic field measurements of B3, B4, and B5 samples which showed a superconducting transition. In contrast, B1 and B2 samples failed to conduct at all, due to their discontinuous layers of Al. Sample B3 has the highest critical field of 1.1 T, in part because of having a thinner Al layer of 8.5 nm.

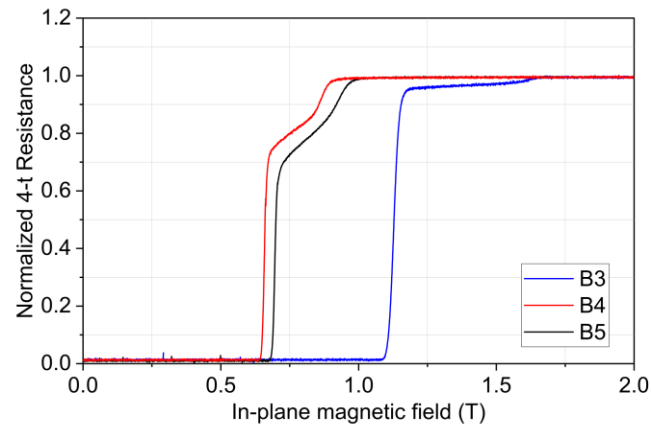

**Figure 6.** In-plane magnetic field measurements for B3, B4, and B5 samples. It shows the relationship between critical magnetic fields and normalized 4-point resistance values at 20 mK.

Aluminum, a type I superconductor, typically undergoes an abrupt transition from superconducting to the normal state with an increasing magnetic field. However, in all tested samples, a second transition for the critical magnetic field is observed, which we attribute to small variations in the Al thickness across the Al layer in each sample. If a film exhibits interwoven patches having two distinct thicknesses interconnected in series along the possible current paths, two transitions will be evident: the thicker patches will transform to a normal state at a lower magnetic field compared to the thinner ones. The resistivity should drop to zero only when there's a continuous superconducting path. The initial drop in resistances seen in **Figure 6**, most likely, is the result of only the thinner patches becoming superconducting, while the entire layer achieves superconductivity only at the still lower magnetic field.

Remarkably, the out-of-plane critical field for thin Al films is very small (about 10 mT) and does not depend on the layer thickness. Even though our samples were intentionally mounted to have a magnetic field along the surface, even small misalignment would introduce a perpendicular component, effectively lowering the measured critical field. For example, if the in-plane critical field of our Al film was 2 T, a misalignment angle as small as 2° would introduce a perpendicular component of the magnetic field sufficient to reduce the measured critical field to below 1 T. Additionally, the residual strain in the Aluminum layers can be another factor affecting the critical magnetic field. We note that absolute values of critical fields alone cannot serve as quality indicators. Importantly, the critical fields reported here for the continuous layers are sufficiently large for pursuits of MZMs quasiparticles.

## D. Discussion

Several comprehensive studies of the deposition of Al on GaAs surfaces[16,20-22] reveal a complex behaviour of Al, strongly dependent on the starting GaAs surface reconstruction. There is a consensus that, for As-rich surface reconstruction, up to three monolayers of Al can be deposited before the transition to 3D islanding is observed. This is likely due to aluminum's preference to form covalent bonds with surface arsenic. We expect similar mechanisms to be active for the case of Al deposition on $In_{0.75}Ga_{0.25}As$ As-rich surface. Since Al-As bonds are stronger than Ga-As or In-As, aluminum may engage not only the excess As on the surface but also displace some of the surface Ga and In atoms from the existing bonds with As. Since reported here Al depositions were performed on stationary wafers, we do not have enough experimental data to speculate about the surface evolution during the growth of initial Al layers on $In_{0.75}Ga_{0.25}As$. However, the available RHEED observations are consistent with surface evolution resembling that reported by Tournet[16] for the Al deposition on GaAs 2×4 surface with the transition to 3D growth at about 3 Al monolayers. At that point, it becomes energetically favorable for arriving Al to grow isolated Al islands rather than attach to the surface. Here, we propose a simple model that qualitatively explains our key finding, namely 3D-2D morphological transition at high Al growth rates: Once the growth enters the stage when the attachment to islands becomes energetically favorable, one expects the density of small "seed" islands to increase with decreasing Al surface mobility. So, for a given flux of arriving Al atoms, the Al islands density should increase with lowering the substrate temperature. On the other hand, at a fixed temperature, the Al islands' density should increase with increasing Al flux, i.e. increasing the target growth rate. Indeed, a higher Al arrival rate will increase the population of mobile Al on the surface, thus promoting Al-Al encounters and leading to the formation of new seed islands. The higher the seed island density, the more likely their coalescence and formation of a continuous Al layer, i.e. 3D-2D morphological transition. For the case of Al deposition in an MBE system, such as described here, we can fix Al flux, but the substrate temperature rises throughout the deposition process, and the rate of this temperature rise increases with the increase in the nominal Al flux. It is reasonable to assume that once the Al seed islands reach some critical size, their density will remain constant throughout the remaining layer deposition. Indeed, since the temperature is continuously increasing and the Al flux is constant, there will be no drive to form new islands. Then more mobile Al atoms will quickly attach to the existing islands, whose density was established at lower temperatures when Al diffusivity was lower.

However, the process is more complex. Indeed, based on the above considerations alone, one would expect that with a given starting substrate temperature, there will be a critical minimum Al flux above which very dense seeding of Al islands would take place, leading in turn to their early coalescence and subsequent 2D growth. Yet, the observation of larger islands with smaller surface density for faster deposition rate, shown in **Figure 3b**, signals the presence of other mechanisms. Indeed, at a growth rate of 0.1 Å/s, a high density of small Al islands is observed (**Figure 3a**), while at a 0.5 Å/s growth rate (**Figure 3b**), a smaller density of larger Al islands is seen. This observation can be explained by the much faster temperature rise rate for the 0.5 Å/s growth rate than the 0.1 Å/s one. Thus, it is plausible that because the 0.1 Å/s sample spent considerably more time at lower temperatures than the 0.5 Å/s sample, it established the Al seed island surface density at a much lower temperature than the latter. Since surface diffusivity depends exponentially on temperature, it may have been the dominating factor setting island density rather than Al arrival rate.

As it is clear from the above, the path to forming a continuous, very thin Al layer of uniform thickness is a complex one but depends primarily on two parameters - the starting substrate temperature and the nominal Al growth rate. Importantly, with a sufficiently fast Al deposition rate, it is possible to grow 10 nm thick uniform Al layers showing a high in-plane critical magnetic field necessary for establishing conditions for forming MZM quasiparticles.

## IV. SUMMARY AND CONCLUSIONS

In this study, we successfully demonstrated the in-situ growth of continuous, 2D 10 nm thick Al films on an $In_{0.75}Ga_{0.25}As$ surface, starting at substrate temperatures above approximately 10℃. We investigated the impact of varying Al growth rates on the formation of Al layers and explored how substrate temperature influences the growth mode of Al. We show a significant variation in substrate temperature evolution during growth, which strongly depends on the Al growth rate. A high growth rate of 3 Å/s was shown to produce very smooth, continuous Al layers with a critical in-plane magnetic field higher than 0.63 T at 20 mK for 10 nm thick layers. In contrast, layers grown at growth rates smaller than 1 Å/s showed 3D Al islands and were not conducting during the low-temperature resistivity measurements. Interestingly, the transition from 3D and 2D layer morphology is abrupt, taking place over an Al flux change smaller than 1%. The results show the path for the growth of high-quality superconducting thin Al layers in a standard configuration MBE system without the need for cryogenic UHV chambers dedicated to in-situ Al deposition.

## ACKNOWLEDGMENTS

*We acknowledge support from funding sources and institutions. This research was made possible partly by the financial aid from the Canada First Research Excellence Fund, specifically the Transformative Quantum Technologies (TQT) program, and the Natural Sciences and Engineering Research Council (NSERC) of Canada. The research was primarily conducted in the QNC-MBE lab. Additionally, significant use was made of the University of Waterloo's QNFCF Facility. The realization of this infrastructure owes much to the substantial contributions from CFREF-TQT, the Canada Foundation for Innovation (CFI), Innovation, Science and Economic Development Canada (ISED), the Ontario Ministry of Research and Innovation, as well as the generous support of Mike and Ophelia Lazaridis. The assistance and resources provided by all these entities are deeply appreciated and have been instrumental in the progression of this work.*

## AUTHOR DECLARATIONS

**Conflicts of Interest**

The authors have no conflicts to disclose.

**Author Contributions** *(if applicable)*

## DATA AVAILABILITY

The data that support the findings of this study are available from the corresponding author upon reasonable request.

# REFERENCES

[1] S. D. Sarma, M. Freedman, and C. Nayak, npj Quantum Information **1**, 1 (2015).

[2] D. Aasen, M. Hell, R. V. Mishmash, A. Higginbotham, J. Danon, M. Leijnse, T. S. Jespersen, J. A. Folk, C. M. Marcus, and K. Flensberg, Phys. Rev. X **6**, 031016 (2016).

[3] C. Wang, X. Li, H. Xu, Z. Li, J. Wang, Z. Yang, Z. Mi, X. Liang, T. Su, and C. Yang, npj Quantum Information **8**, 3 (2022).

[4] P. Krogstrup, N. Ziino, W. Chang, S. Albrecht, M. Madsen, E. Johnson, J. Nygård, C. M. Marcus, and T. Jespersen, Nature Mater. **14**, 400 (2015).

[5] J. Shabani, M. Kjærgaard, H. J. Suominen, Y. Kim, F. Nichele, K. Pakrouski, T. Stankevic, R. M. Lutchyn, P. Krogstrup, and R. Feidenhans, Phys. Rev. B **93**, 155402 (2016).

[6] P. Tedrow and R. Meservey, Phys. Rev. B **25**, 171 (1982).

[7] R. M. Lutchyn, E. P. Bakkers, L. P. Kouwenhoven, P. Krogstrup, C. M. Marcus, and Y. Oreg, Nat. Rev. Mater. **3**, 52 (2018).

[8] C. Thomas, A. Hatke, A. Tuaz, R. Kallaher, T. Wu, T. Wang, R. Diaz, G. Gardner, M. Capano, and M. Manfra, Phys. Rev. Mater. **2**, 104602 (2018).

[9] V. Mourik, K. Zuo, S. M. Frolov, S. R. Plissard, E. P. A. M. Bakkers, and L. P. Kouwenhoven, Science Report **336**, 1003 (2012).

[10] O. Kürtössy, Z. Scherübl, G. Fülöp, I. E. Lukács, T. Kanne, J. Nygård, P. Makk, and S. Csonka, NPJ Quantum Mater. **7**, 88 (2022).

[11] E. Cheah, D. Z. Haxell, R. Schott, P. Zeng, E. Paysen, S. C. ten Kate, M. Coraiola, M. Landstetter, A. B. Zadeh, A. Trampert, M. Sousa, H. Riel, F. Nichele, W. Wegscheider, and F. Krizek, Phys. Rev. Mater. **7**, 073403 (2023).

[12]W. Chang, S. Albrecht, T. Jespersen, F. Kuemmeth, P. Krogstrup, J. Nygård, and C. M. Marcus, Nat. Nanotechnol. **10**, 232 (2015).

[13]S. Frolov, M. Manfra, and J. Sau, Nat. Phys. **16**, 718 (2020).

[14] M. Aghaee, A. Akkala, Z. Alam, R. Ali, A. Alcaraz Ramirez, M. Andrzejczuk, A. E. Antipov et al. (Microsoft Quantum Collaboration), Phys. Rev. B **107**, 245423 (2023).

[15]S. Pilkington and M. Missous, J. Cryst. Growth **196**, 1 (1999).

[16]J. Tournet, D. Gosselink, G.-X. Miao, M. Jaikissoon, D. Langenberg, T. G. McConkey, M. Mariantoni, and Z. R. Wasilewski, Supercond. Sci. Technol. **29**, 064004 (2016).

[17]W. L. Sarney, S. P. Svensson, K. S. Wickramasinghe, J. Yuan, and J. Shabani, J. Vac. Sci. Technol. B **36**, 062903 (2018).

[18]S. R. Johnson, "Optical bandgap thermometry in molecular beam epitaxy," PhD Dissertation, University of British Columbia, Vancouver, 1995.

[19]Z. R. Wasilewski, G. C. Aers, A. J. SpringThorpe, and C. J. Miner, Journal of Vacuum Science & Technology B (Microelectronics Processing and Phenomena) **9**, 120 (1991).

[20]S. Donner, R. Blumenthal, J. Herman, R. Trehan, E. Furman, and N. Winograd, Appl. Phys. Lett. **55**, 1753 (1989).

[21]G. Landgren, R. Ludeke, and C. Serrano, J. Cryst. Growth **60**, 393 (1982).

[22]J. Massies and N. T. Linh, Surf. Sci. **114**, 147 (1982).